\documentstyle[procl]{article}

\input{psfig}

\def\Journal#1#2#3#4{{#1} {\bf #2}, #3 (#4)}

\def\NPB{{\em Nucl. Phys.} B}

\def\PLB{{\em Phys. Lett.}  B}

\def\PRD{{\em Phys. Rev.} D}

\def\ZPC{{\em Z. Phys.} C}

\def\be{\begin{equation}}

\def\ee{\end{equation}}

\def\bea{\begin{eqnarray}}

\def\eea{\end{eqnarray}}

\begin{document}

\title{THE LONGITUDINAL STRUCTURE FUNCTION $F_L$
AT SMALL $X$}

\author{ A.V. KOTIKOV }

\address{Laboratoire de Physique Theorique ENSLAPP\\ LAPP, B.P. 100,
F-74941, Annecy-le-Vieux Cedex, France\\ and \\
Particle Physics Laboratory, JINR, Dubna, Russia}

\author{ G. PARENTE }

\address{Departamento de F\'\i sica de Part\'\i culas\\
Universidade de Santiago de Compostela\\
15706 Santiago de Compostela, Spain}

%%%%%%%%%%%%%%%%%%%%%%%%%%%%%%%%%%%%%%%%%%%%%%%%%%%%%%%%%%%%%%


%%%%%%%%%%%%%%%%%%%%%%%%%%%%%%%%%%%%%%%%%%%%%%%%%%%%%%%%%%%%%%

\maketitle\abstracts{We present 
a set of formulae to extract the longitudinal
deep inelastic  structure function $F_L$
from the transverse structure function $F_2$ and
its derivative $dF_2/dlnQ^2$ at small $x$. 
Using $F_2$ HERA data we obtain
$F_L$
in the range $10^{-4} \leq x \leq  10^{-2}$ at $Q^2=20$ GeV$^2$.}

%\section{Guidelines}

%\subsection{Producing the Hard Copy}\label{subsec:prod}

We present here the values of
$F_L(x,Q^2)$ at
small $x$, extracted from experimental HERA data \cite{F2H1,F2ZEUS}
using the method 
of replacement of the Mellin convolution by ordinary products \cite{2}.
By analogy with the case of the gluon distribution function (see
\cite{KOPA} and its references)
it is possible to obtain the relation between $F_L(x,Q^2)$, $F_2(x,Q^2)$
and $dF(x,Q^2)/dlnQ^2$ at small $x$. Thus, the small $x$ behaviour of
the SF $F_L(x,Q^2)$ can be extracted directly from the measured values
of $F_2(x,Q^2)$ and its derivative without a cumbersome
procedure \cite{1.5}. 
These extracted values of $F_L$ may be well considered as
{\it new
small $x$ ``experimental data''}.
When experimental data for $F_L$ at small $x$ become
available with a good
accuracy \footnote{In the time of preparing this article, the H1
  collaboration presented \cite{H1FL} the first (preliminary) measurement
of $F_L$ at small $x$}, a violation of the relation will be
an indication
of the
importance of other effects as higher twist contribution and/or
of non-perturbative QCD dynamics at small $x$.

We follow the notation of our previous work in ref. \cite{KOPA}. 
The singlet quark
$s(x,Q^2_0)$ and gluon $g(x,Q^2_0)$ parton distribution 
functions (PDF) \footnote{We use PDF
%parton distribution functions 
multiplied by $x$
and neglect the nonsinglet quark distribution at small $x$.}
at some $Q^2_0$ are parameterized by
  (see, for example, \cite{4}) \begin{eqnarray} 
p(x,Q^2_0) & = & A_p
x^{-\delta} (1-x)^{\nu_p} (1+\epsilon_p \sqrt{x} + \gamma_p x)
~~~~(\mbox{hereafter } p=s,g)
\label{1} \end{eqnarray}

Futher, we restrict the analysis to the case of large $\delta $ values
(i.e. $x^{-\delta} \gg 1$ ) which correspond to BFKL pomeron which is
supported by HERA data. The more complete analysis may be found in
\cite{KOPAFL}, where we took into account also the case $\delta \sim
0$ corresponding to the standard pomeron.%\\ 

 Assuming the {\it Regge-like behaviour} for the gluon distribution and
 $F_2(x,Q^2)$: 
%at $x^{-\delta} \gg 1$:
$$g(x,Q^2)  =  x^{-\delta} \tilde g(x,Q^2),~~ 
F_2(x,Q^2)  =  x^{-\delta} \tilde s(x,Q^2), $$
 we obtain the
following equation for the $Q^2$ derivative of
the SF $F_2$:
 \begin{eqnarray}  \frac{dF_2(x,Q^2)}{dlnQ^2}  &=&
-\frac{1}{2} 
%\delta_s 
x^{-\delta} \sum_{p=s,g}
\Bigl( 
 r^{1+\delta}_{sp}(\alpha) ~\tilde p(0,Q^2) +
 r^{\delta}_{sp}(\alpha)~ x \tilde p'(0,Q^2)  +
 O(x^{2}) \Bigr) \nonumber \\ 
F_L(x,Q^2)  &=&
%\delta_s 
x^{-\delta} \sum_{p=s,g}
\Bigl( 
 r^{1+\delta}_{Lp}(\alpha) ~\tilde p(0,Q^2) +
 r^{\delta}_{Lp}(\alpha)~ x \tilde p'(0,Q^2)  +
 O(x^{2}) \Bigr) , 
\label{2.1} \end{eqnarray}
 where $r^{\eta}_{sp}(\alpha) $ and $r^{\eta}_{Lp}(\alpha) $
 are the combinations (see \cite{KOPAFL})
of the anomalous dimensions of Wilson operators 
and Wilson 
coefficients
of the $\eta$
"moment"  (i.e., the corresponding variables extended
from integer values of argument to non-integer ones).
With accuracy of $O(x^{2-\delta}, \alpha x^{1-\delta})$ 
(see \cite{KOPA,KOPAFL}),  
we have for Eq.(\ref{2.1}) 
\begin{eqnarray} 
F_L(x, Q^2)  &=& -
\xi ^{\delta}
  \biggl[ 2 \frac{ r^{1+\delta}_{Lg}}{ r^{1+\delta}_{sg}} 
%\cdot
\frac{d F_2(x \xi , Q^2)}{dlnQ^2}
+ \biggl( r^{1+\delta}_{Ls} - \frac{ r^{1+\delta}_{Lg}}{ r^{1+\delta}_{sg}}
r^{1+\delta}_{ss} \biggr) F_2(x \xi ,Q^2)  \nonumber \\ 
 &+&~  
O(x^{2-\delta},\alpha x^{1-\delta}) \biggr] 
\label{8}\end{eqnarray}
Using NLO approximation of $r^{1+\delta}_{sp}$ and $r^{1+\delta}_{Lp}$
for concrete values of $\delta = 0.5$ 
%(see also \cite{KOPA}) 
and $\delta = 0.3$ we obtain (for f=4 and $\overline{MS}$ scheme):
\begin{eqnarray} 
%\z ~\mbox{ if }~ \delta =0.5 \nonumber \\ \z
%
F_L(x, Q^2)  &=& \frac{0.87}{ 1 + 22.9 \alpha} \Biggl[
  \frac{d F_2(0.70 x , Q^2)}{dlnQ^2} + 
4.17 \alpha F_2(0.70 x,Q^2) \Biggr] \nonumber \\ 
%\z ~~~~~~~ ~~~~~~~~
&+&~O(\alpha^2,x^{2-\delta},\alpha x^{1-\delta}) 
~~~~~\mbox{ if }~ \delta =0.5 
\label{10.1} 
\\  \nonumber 
\\ 
%\z~\mbox{ if }~ \delta =0.3 \nonumber \\ \z
F_L(x, Q^2)  &=& \frac{0.84}{1 + 59.3 \alpha } \Biggl[ 
 \frac{d F_2(0.48 x, Q^2)}{dlnQ^2} + 3.59 \alpha 
F_2(0.48 x , Q^2) \Biggr]    \nonumber \\ 
%\z ~~~~~~~ ~~~~~~~~
&+&~ O(\alpha^2 , x^{2-\delta} ,\alpha x^{1-\delta}) 
 ~~~~~\mbox{ if }~ \delta =0.3        
\label{10.3}
%\\ \z
%F_L(x, Q^2)  = \frac{1.05}{1 + 59.3 \alpha } 
% \frac{d F_2(x, Q^2)}{dlnQ^2} + 3.76 \alpha 
%F_2(x , Q^2) + O(\alpha^2 \!\!, x^{1-\delta}) 
%\!         
%\label{10.4} 
\end{eqnarray}

%{\bf 2.} 
With the help of Eq. (\ref{10.3})
% ??? may be better to use Eq. (\ref{10.3})??? 
we have extracted the longitudinal SF $F_L(x,Q^2)$
from HERA data, using the slopes dF$_2$/dlnQ$^2$
of H1 \cite{F2H1} and ZEUS \cite{F2ZEUS} HERA data. 
Fig. 1 shows the extracted values of the longitudinal SF
and the QCD prediction from set MRS(G). The agreement
is excellent.
There is also a relative good agreement with a recent 
experimental H1 point for $F_L$ \cite{H1FL}, if one takes into account the
systematic error.

%{\bf 3.}
In summary, we have presented  Eqs. (\ref{8})-(\ref{10.3}) for the
extraction of the longitudinal SF $F_L$ at small $x$ from the
$F_2$ and its
$Q^2$ derivative. These equations provide the possibility of the non-direct
determination of $F_L$. This is important since the direct
extraction of $F_L$ from experimental data is a cumbersome
procedure \cite{1.5}.
Moreover, the fulfillment of  Eqs. (\ref{8})-(\ref{10.3}) in DIS
experimental data can be used as a cross-check of perturbative QCD
at small values of
$x$.

\begin{figure}[hbt]
\centering
\vspace*{-2.cm}
%\vskip 2.5cm
%\rule{5cm}{0.2mm}\hfill\rule{5cm}{0.2mm}
%\vskip 2.5cm
\psfig{figure=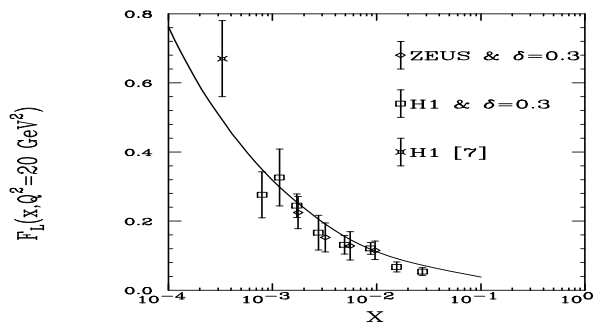,height=8.5cm,width=12cm}
\vspace*{-2.0cm}
%\rule{5cm}{0.2mm}\hfill\rule{5cm}{0.2mm}

\caption{The extracted longitudinal SF $F_L$ (see text for details)}

%Solid line is the calculation from set MRS(G) \cite{4}
%using O($\alpha_s^2$) corrections. It is also shown a preliminary
%H1 data point.}
%\label{fig:radish}

\end{figure}

\vspace*{-0.6cm}

\section*{Acknowledgments}
This work was supported in part by CICYT and by Xunta de Galicia.
We are grateful to A. Bodek and M. Klein for discussions.

\vspace*{-0.3cm}

%
%                 REFERENCIAS
%

%
%

\end{document}